\documentclass[aps,twocolumn, 10pt]{revtex4}

\makeatletter
\def\@dotsep{4.5}
\makeatother

\usepackage{mathrsfs}
\usepackage{amssymb}
\usepackage{color}
\usepackage{latexsym}
\usepackage{epsf}
\usepackage{epsfig,bm}
\begin{document}
\title{Energy Landscape, Anti-Plasticization and Polydispersity Induced
Crossover of Heterogeneity in Supercooled Polydisperse Liquids}
\author{Sneha Elizabeth Abraham}
\author{Sarika Maitra Bhattacharrya}
\author{Biman Bagchi\footnote[1]{Electronic mail: bbagchi@sscu.iisc.ernet.in}}
\affiliation{Solid State and Structural Chemistry Unit, Indian Institute of
Science, Bangalore 560 012, India}

\begin{abstract}
Polydispersity is found to have a significant effect on the potential energy
landscape; the average inherent structure energy 
decreases with polydispersity. Increasing polydispersity at fixed volume
fraction decreases the glass transition temperature and the fragility of glass formation
analogous to the {\em antiplasticization} seen in some polymeric melts. An interesting
temperature dependent crossover of heterogeneity with polydispersity is observed at
low temperature due to the faster build-up of dynamic heterogeneity at lower
polydispersity.

\end{abstract}
\maketitle

Polydispersity is ubiquitous in nature. It is present in clays, minerals, paint
pigments, metal and ceramic powders, food preservatives and in simple homogeneous
liquids. It is common in synthetic colloids, which 
frequently exhibit considerable size polydispersity \cite{pusey} and is also found
in industrially produced polymers, which contain macromolecules with a
range of chain length. Polydispersity has significant effects on both the structure and
dynamics of the system. Experiments \cite{phan} and
simulations \cite{rastogi,sear} on colloidal
systems show that increasing polydispersity, at a constant volume fraction, lowers
structural correlations, pressure, energy and viscosity. 
Polydisperse colloidal systems are known to be excellent glass formers. William et al
\cite{william} suggest that colloidal glass formation results from a small degree of particle
polydispersity. Crystal nucleation in a polydisperse colloid is suppressed due to the
increase of the surface free energy \cite{frenkel}. Studies by several groups \cite{pinaki} have shown
that the glass becomes the equilibrium phase beyond a terminal value of polydispersity.

Despite being natural glass formers, relationships between polydispersity,
fragility, energy landscape and heterogeneous dynamics have not been adequately
explored in these systems. Because these systems exist in the glassy phase over a
wide range of polydispersity, they offer opportunity to test many of the theories and
ideas developed in this area in recent years. We find that polydispersity introduces
several unique features to the dynamics of these systems not present in the binary
systems usually employed to study dynamical features in supercooled liquids and
glasses.

In this work we particularly investigate how polydispersity influences the
potential energy landscape, fragility and heterogeneous dynamics of polydisperse
Lennard-Jones (LJ) systems in supercooled regime near the glass transition \cite{murarka}. The
polydispersity in size is introduced by random sampling from a Gaussian distribution
of particle diameters, $\sigma$. 
The standard deviation $\delta$
of the distribution divided by its mean $ \overline \sigma$ gives a dimensionless
parameter, the polydispersity index $S = \frac {\delta} {\overline \sigma}$.
The  mass $m_{i}$ of particle $i$ is scaled by its diameter
as $m_{i} = \overline m(\frac{\sigma_{i}}{\overline \sigma})^3$.⎝ ⎠
Micro canonical (NVE)
ensemble MD simulations are carried out at a fixed volume fraction, $\phi$ on a system of
$N=864$ particles of mean diameter $\sigma=1.0$ and mean mass $m=1.0$ for $S=0.10$,
$0.15$ and $0.20$ at $\phi = 0.52$ and $S=0.10$ and $0.20$ at $\phi=0.54$.
All quantities in this
study are given in reduced units (length in units of $\sigma$, temperature in units of
$\frac{\epsilon}{k_{B}}$ and time in units of
$\tau=( \frac{\overline m  \overline \sigma^{2}}{\epsilon } ) ^{\frac{1}{2}}$).
The LJ interaction parameter $\epsilon$ is assumed to have the same
value for all particle pairs.

\begin{figure}
 \begin{center}
 \epsfig{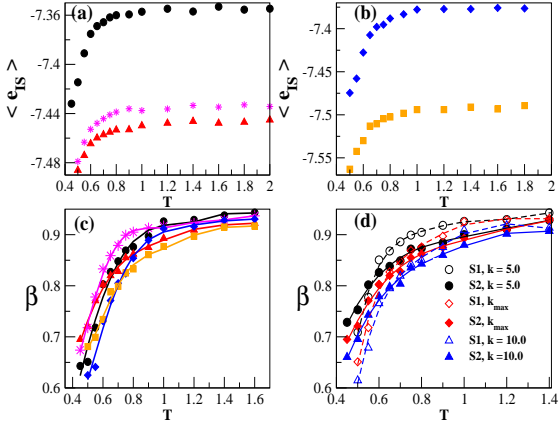}
 \caption{(a) and (b) Temperature dependence of 
the average inherent structure energy, $\langle e_{IS} \rangle$. 
For Fig \ref{eis-beta} (a), (b) \& (c), filled circles, stars and triangles are for $S=0.10$, $S=0.15$ \& $S=0.20$
at $\phi=0.52$ and filled diamonds and squares are for $S=0.10$ \& $S=0.20$
at $\phi=0.54$, respectively. (c) The stretched exponent $\beta$ vs. T obtained by
fitting KWW equation to self-intermediate scattering function, $F_{s}( k_{max}, t)$
where $k_{max}\sim 7.0$. 
The lines are 
guide to the eye. Comparison between (a)/(b) and (c) shows that the fall of
$\langle e_{IS} \rangle$ corresponds to the onset of non-exponential relaxation in
$F_{s}( k_{max}, t)$. (d) $\beta$ vs. T from $F_{s}( k, t)$ for different k values. Data shown for 
$S=0.1(S1)$ \& $S=0.2(S2)$ at $\phi=0.52$. $S=0.15$ omitted for clarity.  }
 \label{eis-beta}
 \end{center}
 \end{figure}

At large supercooling the system settles into glassy phase. We first analyze the
system from the perspective of potential energy landscape (PEL), which has emerged
as an important tool in the study of glass forming liquids \cite{sastry, stillinger, wales}.
Fig \ref{eis-beta}(a) and (b) show
the variation of the average inherent structure energy ($\langle e_{IS} \rangle$)
with temperature ($T$) at
both the volume fractions studied. The value of $\langle e_{IS} \rangle$ remains fairly
insensitive to the
variation in $T$ at high T before it starts to fall with $T$ (around $T \sim 1.0$).
It
has been established earlier in the context of the binary mixtures \cite{sastry, dwc} that
{\em the start
of fall in $\langle e_{IS} \rangle$ coincides with the onset of non-exponential
relaxation in the time
correlation functions of the system}. We show in Fig \ref{eis-beta}(c) that this correlation
continues to hold in polydisperse systems.
The fall of $\langle e_{IS} \rangle$ with $T$ is consistent with
the Gaussian landscape model.
\begin{figure}
 \begin{center}
 \epsfig{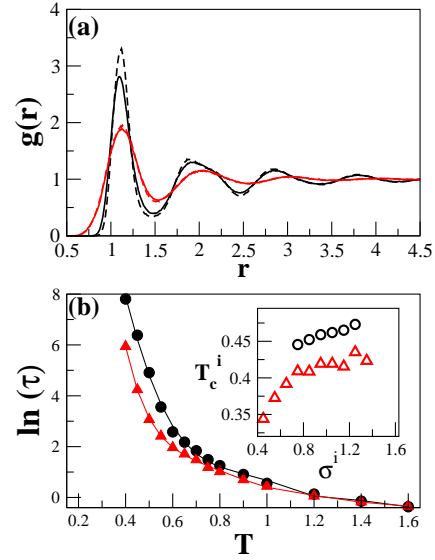}
 \caption{ (a) Average radial distribution functions (rdf) for the parent liquid (solid line)
and the inherent structure (dashed line) for $S=0.10$ (black) and $S=0.20$ (red)
systems at $T=0.50$ and $\phi=0.52$. (b) Relaxation time $\tau$ from KWW fit to
$F_{s}( k_{max}, t)$ for $S=0.10$ (filled circles) and $S=0.20$ (filled triangles) at $\phi=0.52$.[Inset: Critical temperature $T_{c}^{i}$  for particles
of different sizes $\sigma^{i}$ obtained from the MCT equation, $D^{i}\sim(T-T_{c}^{i})^{\gamma}$ for $S=0.10$ (open circles) and $S=0.20$ (open triangles) at $\phi=0.52$.] }
 \label{rdf}
 \end{center}
 \end{figure}

The average inherent structure energy decreases with polydispersity (Fig
\ref{eis-beta}(a) and (b)), which indicates that the packing is more efficient at higher S. In
Fig \ref{rdf} we plot the inherent structure (IS) and the parent liquid radial distribution
functions (rdf). At $S=0.20$ there is hardly any difference between the rdf of the
parent liquid and the IS. The coordination number, $N_{c}$ at $S=0.10$ and $S=0.20$
obtained from the IS rdf are $13.1$ and $14.6$, respectively. This shows that packing is
more efficient at higher S and one would expect a slowing down of
dynamics at higher S. 
Instead, we find that similar to colloidal hard spheres polydisperse LJ systems also show
a speed up of relaxation with S. 
The presence of smaller particles at higher S provides some sort of lubrication
\cite{lionberger,wil-megen}, which speeds up the dynamics of the whole system.
A plot of the Mode Coupling Theory (MCT)\cite{sarika} critical temperature $T_{c}^{i}$
for particles
of different sizes $\sigma^{i}$ (inset of Fig \ref{rdf}(b)) shows
that the $T_{c}^{i}$ for the largest-sized particles in S=0.20 system is smaller
than the smallest-sized particles in S=0.10 system. This tells us that not only the smaller
particles in $S=0.20$ system but the whole system has a faster relaxation.
The rate of growth of relaxation time upon lowering of T
decreases with S (Fig \ref{rdf}(b)). Hence
as the system is cooled, vitrification is expected to occur at a lower T 
for the system at higher S. This should lead to a lowering of the
glass transition temperature with S.  

 \begin{figure}
 \begin{center}
 \epsfig{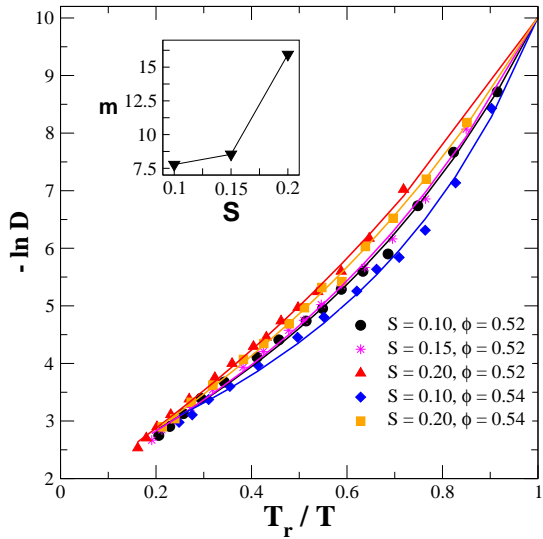}
 \caption{Angell-like fragility plot at different S for the two $\phi$
studied. The thick
lines are VFT fit to the diffusivity data, $D=D_{0}exp(\frac{E_{D}}{T-T_{0}})$.
The reference temperature $T_{r}$ is chosen such that $D(T_{r})=4.5 \times 10^{-5}$.
The VFT extrapolation is used
to locate $T_{r}$. The plot shows that fragility decreases with $S$ and that for a given $S$
fragility increases with increase in  $\phi$. [Inset: Strength parameter $m$
(where $m=\frac{E_{D}}{T_{0}}$ \cite{Angell}) obtained from VFT fit as a function of $S$ at $\phi=0.52$.]
 }
 \label{fragility}
 \end{center}
 \end{figure}

Fragility is a term being used to characterize and quantify the non-Arrhenius transport
behavior in glass-forming liquids as they approach glass transition\cite{Angell}. To study the
effect of polydispersity on fragility, we plot the diffusion coefficients in an
Angell-like fragility plot in Fig \ref{fragility}. The plot clearly shows that {\em increasing polydispersity} at
fixed volume fraction {\em reduces the fragility} of the liquid so that the system is a
stronger glass former at higher polydispersity. This effect is analogous to the
{\em antiplasticization} that has been observed in polymer melts
\cite{plast1}. 
PEL analysis shows that the antiplasticized system has smaller barriers to overcome in
order to explore the configuration space \cite{plast2}. In the rest of the paper we explore the correlations between fragility and non-exponential relaxation/heterogeneous dynamics.

Fragility is usually correlated to the stretch exponent $\beta$ which is found
to be valid for many materials \cite{ngai-niss}. 
From PEL perspective, fragile liquids
display a proliferation of well-separated basins which result
in a broad spectrum of relaxation times leading to stretched exponential dynamics
\cite{stillinger}.
The correlation is also consistent within the
framework of coupling model (CM) \cite{CM} according to which the strength of the
intermolecular coupling is given by ($1-\beta$). 
The rate of growth of intermolecular coupling with
decrease in T is a measure of fragility which according to CM would
depend on the rate of fall of $\beta$ with T.
We indeed find that as S increases (fragility decreases) the rate of fall of
$\beta$ with T decreases (Fig \ref{eis-beta}(c)).
However, if we look only at the $\beta$ values and not its T-dependence   
we find that at  high T, stretching is anti-correlated with fragility
whereas {\em at low T, we get the
reverse scenario where the
stretching is correlated with fragility}.
This leads to a cross-over of the $\beta$ values for different S
at intermediate T as shown in Fig \ref{eis-beta}(c). 
The $\beta$ values in Fig \ref{eis-beta}(c) are obtained by
KWW fit to $F_{s}( k_{max}, t)$. However, these cross-overs are independent
of $k$ values as shown in 
Fig \ref{eis-beta}(d).  
The interplay between
the T-independent {\em intrinsic heterogeneity} 
(due to the particle
size and mass distribution) and the dynamic heterogeneity
which builds up at low $T$ seems to be the 
microscopic origin of the anti-correlation
between fragility and stretching at high T
and the observed crossover at intermediate T.
\begin{figure}
 \begin{center}
 \epsfig{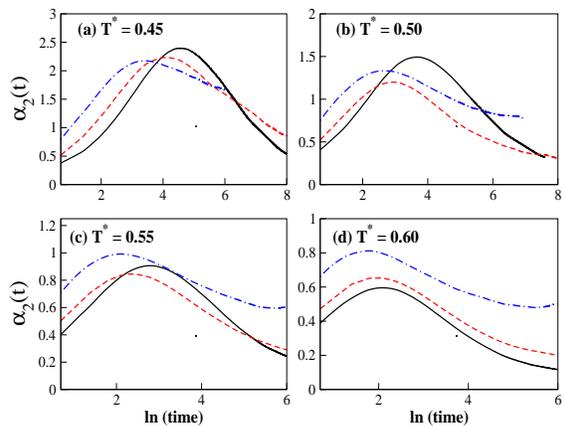}
 \caption{The non-Gaussian parameter, $\alpha_{2}(t)$ for $S=0.10$ (solid line),$S=0.15$
(dashed line) and $S=0.20$ (dot-dashed line) at four different T depicting the crossovers between different $S$.
Data is shown for $\phi = 0.52$. }
 \label{ngp}
 \end{center}
 \end{figure}

To investigate this point in further
details, we study the non-Gaussian parameter,
$\alpha_{2}(t)$ which also shows a correlation with fragility for
most materials \cite{sokolov}. 
The non-zero values
of $\alpha_{2}(t)$ in a monodisperse system is purely
due to the presence of dynamic heterogeneity whereas
in polydisperse system, in
addition to dynamic heterogeneity, there is an ‘intrinsic heterogeneity’ due to particle
size and mass distribution which is present at all $T$. 
Thus for the latter, $\alpha_{2}(t)$ reflects a coupled effect of
both these heterogeneities.
As seen in Fig \ref{ngp}, for a polydisperse system $\alpha_{2}(t)$ is nonzero
both in the short time limit
(due to the mass distribution \cite{poole}) and in the
long time limit (due to the
spread in diffusion coefficients with particle size and mass).
At high $T$, the non-zero value of $\alpha_{2}(t)$ is predominantly
due to the intrinsic heterogeneity and thus increases with S (Fig \ref{ngp}(d)).
As $T$ is lowered, the effects of dynamic heterogeneity
starts to dominate, as was shown by the onset of connected clusters of fast moving
particles \cite{sear, weeks} whose
size increases as one approaches glass transition.
Since the relaxation time increases with decrease
of S (Fig. \ref{rdf}(b)), there is a faster build-up of dynamic heterogeneity at lower
S which leads to the observed crossovers (Fig \ref{ngp}(c)\&(b)) in the values of
$\alpha_{2}(t)$ between different S (similar to that observed for $\beta$ in
Fig \ref{eis-beta}(c)).
Hence at low $T$, one gets the scenario where $\alpha_{2}(t)$ decreases with
polydispersity (Fig \ref{ngp}(d)). 
Since fragility decreases with S, these crossovers in $\beta$ and $\alpha_{2}(t)$ {\em 
would mean 
that fragility is correlated only to the dynamic heterogeneity and not to the intrinsic
heterogeneity in the system}.

When the polydispersity is increased at constant volume, we get results that
are opposite to that obtained from constant volume fraction studies. We find that the
dynamics slows down with increase in polydispersity \cite{poole}. This is because at constant
volume as polydispersity increases, the packing fraction increases \cite{cai} and hence we
find a coupled effect of polydispersity and density.

Our results show that at constant volume fraction, although the increase of
polydispersity leads to a more efficient packing, the dynamics become faster due to
the lubrication effect.
Fragility decreases with polydispersity 
and is found to be
correlated only to the rate of
growth of dynamic heterogeneity
and not to the intrinsic
molecular heterogeneity in the system. These results reveal that the rich dynamics of
the polydisperse system can lead to new relaxations mechanisms that deserve further
study.

We thank Dr. S. Sastry and Dr. D. Chakrabarti for discussions. This work was supported in
parts by grants from DST, India. S. E. Abraham
acknowledges the CSIR (India) for financial support.

\end{document}